\documentstyle[twocolumn,aps,prl,epsf]{revtex} 
\begin {document} 
\draft 
 
\wideabs{ 
\title{\bf Zero-temperature equation of state of quasi-one dimensional
H$_{\bf 2}$ 
}   
\author{M.C. Gordillo, J. Boronat and J. Casulleras}  
\address{ 
Departament de F\'{\i}sica i Enginyeria Nuclear, Campus Nord B4-B5,  
Universitat Polit\`ecnica de Catalunya. E-08034 Barcelona, Spain}  
\date{\today} 
 
\maketitle 
\begin{abstract} 
We have studied molecular hydrogen in a pure 1D geometry and inside a narrow
carbon nanotube by means of the diffusion Monte
Carlo method. The one-dimensionality of H$_2$ in the nanotube is well 
maintained in a large density range, this system being closer to an ideal 1D
fluid than liquid $^4$He in the same setup.  H$_2$ shares with 
$^4$He the existence of a stable liquid phase and a quasi-continuous 
liquid-solid transition at very high linear densities.  
\end{abstract} 
 
} 

The experimental finding in 1991 \cite{iji} of carbon nanotubes opened 
brand new possibilities both in technology and in fundamental physics.  
The nanoscale of these new materials has led to the
discovery of novel mechanical, chemical and electrical properties
\cite{aja} that suggest exciting new technological applications. One
of the more relevant features of these new materials is their 
large adsorption capability compared with that of a graphite planar substrate.
In fact, the 
enhanced Van der Waals interaction when particles are adsorbed inside 
a single nanotube, or in the interstitial sites of single wall carbon nanotube
(SWCN) bundles, may allow the density of the adsorbed substrate to be high
enough for the bundles to serve as gas storage devices. Special interest
exists in the physisorption of hydrogen \cite{dil,dar,stan1,rze,wan} in the
quest for a fuel cell efficient enough to be
used as a pollution-free energy carrier. It has been recently suggested that
SWCN with diameters of the order of a nanometer can be the best candidates
to approach the pursued level of packing \cite{dil}.   

From a more fundamental point of view, the strong confinement of particles
adsorbed in the carbon channels of a 
SWCN bundle, with diameters ranging form 7 to 40 \AA\ and an width to length
ratio of
$\sim$ 1000, offers the possibility of an experimental realization of a 
quasi-one dimensional 
system. Moreover, if the temperature is low enough one is dealing with a 
unique opportunity of studying a nearly one-dimensional quantum fluid. In a
recent experiment, Teizel {\em et al.} \cite{tei} have unambiguously observed the quasi-one
dimensional behavior of $^4$He adsorbed in a SWCN bundle by measuring its 
desorption rate. On the other hand, theoretical studies in the limit of
zero temperature and  strictly one dimension have proved the existence of
a liquid state with a binding energy in the milli-Kelvin scale
\cite{kro,gor,mor}. Using the diffusion 
Monte Carlo (DMC) method, we also compared the ideal 1D geometry
system
with a realistic situation in which $^4$He is adsorbed in a narrow nanotube
\cite{gor}. 
The aim of the present work is to extend the theoretical study to the appealing
case of H$_2$ that, besides its technological relevance, might offer the
existence of a homogeneous liquid phase at zero temperature. It is worth 
noticing that both bulk and two-dimensional H$_2$, are solid in this temperature
limit. Liquid phases have only been observed in theoretical calculations of
small clusters \cite{sin} and in 2D geometries with localized alkaline 
impurities \cite{gor2}.  

We have studied molecular hydrogen at zero temperature in a 
one-dimensional (1D) array and inside a single walled carbon nanotube (T)
of radius $R$ = 3.42 \AA, which corresponds to a (5,5) armchair tube
\cite{ham}. The
technique used is the DMC method, which has become in the
last decades one of the most efficient theoretical tools, from the microscopic
point of view, to deal with  quantum fluids. The DMC method solves the 
$N$-body imaginary-time Schr\"{o}dinger equation 
\begin{equation}
-\frac{\partial f({\bf R},t)}{\partial t} = -D \nabla^2_R f({\bf R},t) + 
D \nabla_R \cdot ({\bf F}  \, f({\bf R},t)) +
(E_L({\bf R}) -E) f({\bf R},t)   \ ,
\end{equation}
in which $f({\bf R},t)$ = $\psi({\bf R})\Psi({\bf R},t)$, $D = \hbar^2/2m$, 
$E_L({\bf R}) = \psi({\bf R})^{-1} H \psi({\bf R})$ is the local energy, and  
${\bf F}({\bf R}) = 2 \, \psi({\bf R})^{-1}
\nabla \psi({\bf R})$ is the so-called drift force. The time-independent wave
function $\psi({\bf R})$ acts as an important sampling function and is chosen
as a trial model usually adjusted at a variational Monte Carlo (VMC) level.
In bosonic systems like the present one the DMC method provides exact results
apart from statistical uncertainties. More specific details of DMC may be
drawn from Refs. \cite{bor,les}.

A relevant issue in a microscopic study is the nature of the interspecies 
interaction. We have considered the H$_2$ molecules interacting via the isotropic
semiempirical potential from Silvera and Goldman (SG) \cite{sil1} that has been 
extensively used in path integral Monte Carlo (PIMC) and DMC calculations 
of bulk \cite{chen}, clusters \cite{sin} and H$_2$ films \cite{wag}. The SG 
is a pair potential that  incorporates to some extent
the effect of  three-body interactions by means of an effective two-body 
term of the form $C_9/r^9$. On the other hand, the isotropy of the potential 
is well justified if one considers that at very low temperatures almost all 
the H$_2$ molecules are para-hydrogen species, i.e., they are in the J=0
rotational state.  In the simulations of H$_2$ inside a nanotube, we consider
a cylindrically symmetric potential as suggested by Stan and Cole
\cite{stan2}. In that 
simplified model, the interactions between C atoms and H$_2$ molecules are
axially averaged out resulting in a potential which only depends on the 
distance to the center of the tube. It has been proved \cite{stan1} that the differences
between that smoothed potential and a potential which is built up as an 
explicit sum of individual C-H$_2$ interactions are not significant and surely
smaller than the relative uncertainty in the ($\sigma$,$\epsilon$) 
Lennard-Jones parameters. Considering $\sigma$ = 2.97 \AA\ and $\epsilon$ =
42.8 K, the symmetric potential felt by a H$_2$ molecule in a (5,5) tube has a
depth of 42 $\epsilon$, three times larger that the attraction of the same 
molecule in a flat graphitic surface. 

The use of importance sampling in DMC requires the introduction of a trial
wave function that guides the diffusion process to relevant regions of the
walkers phase space. In the 1D system, we consider
\begin{equation}
\Psi^{\rm 1D}({\bf R}) = \Psi_{\rm J}({\bf R})  \ ,
\end{equation}
with $\Psi_{\rm J}({\bf R}) = \prod_{i<j} \exp \left[ -\frac{1}{2} \left
(\frac{b}{r_{ij}} \right)^5
\right]$ a Jastrow wave function with a McMillan two-body correlation factor.
Inside the nanotube, H$_2$ molecules interact with the walls of the cylinder
and therefore we have added an additional one-body term 
\begin{equation}
\Psi^{\rm T}({\bf R}) = \Psi_{\rm J}({\bf R}) \Psi_{\rm c}({\bf R}) 
\end{equation}
with $\Psi_{\rm c}({\bf R}) = \prod_i^N \exp(-c \, r_i^2)$, $r_i$ being the
radial distance of the particle to the center, to avoid the hard core of
the
H$_2$-nanotube interaction. 

Theoretical calculations of 1D $^4$He agree in predicting a liquid-solid phase
transition at high linear densities \cite{kro,gor,mor}. This transition which is 
only possible at absolute zero temperature looks like a nearly continuous one 
without a 
measurable difference between the melting and freezing densities. Following the
same procedure that in our previous work in helium, we have explored the
existence of such a transition in H$_2$. In this ordered phase, we modify
the trial wave function in both the 1D system and inside the nanotube by
multiplying them by a $z$-localized factor $\Psi_{\rm s}({\bf R}) = 
\prod_i^N \exp( -a(z_i- z_{is})^2)$. The sites $z_{is}$ are equally-spaced points
in both the 1D line and the axial direction of the nanotube. 

The variational parameters $a$, $b$ and $c$ have been optimized by means
of VMC calculations. In the liquid phase and near the equilibrium density
$b$= 3.759 \AA, with a  slight increase with  density (at $\lambda$ = 
0.277 \AA$^{-1}$, $b$ = 3.789 \AA), whereas $c$ = 4.908 \AA$^{-2}$ is kept
fixed for all $\lambda$ values. In the solid phase, $b$ = 3.404  \AA, $a$ = 
0.799 \AA$^{-2}$ and $c$ = 5.136 \AA$^{-2}$, with a negligible 
$\lambda$ dependence in the region analyzed.  

The possible existence of a liquid-solid phase transition at high linear
density has been studied in both 1D and inside a narrow nanotube. In Table I,
results for the energy per particle of both systems are reported for the liquid ($a$ = 0) and solid ($a \ne$ 0) phases.  The comparison between the energies
of both phases at the same density shows that their difference  changes  
sign in going from $\lambda$ = 0.312 \AA$^{-1}$ to $\lambda$ =
0.304 \AA$^{-1}$ in 1D and from $\lambda$ = 0.320 \AA$^{-1}$ to 
$\lambda$ = 0.312 \AA$^{-1}$ in the tube. Above these densities,
 the system prefers to be
localized in a solid-like structure with a difference $|E(s) - E(l)|$ that
increases with $\lambda$. When the density decreases 
the liquid phase is energetically preferred and again the
size of the difference $|E(s) - E(l)|$ increases when $\lambda$ diminishes. 
The density value at which this difference becomes zero is estimated to be 
$\lambda = 0.309$ \AA$^{-1}$ in 1D and $\lambda = 0.315$ \AA$^{-1}$ in the
tube, being not possible to distinguish between 
freezing and melting densities. As previously studied in $^4$He \cite{gor} 
it appears to be a nearly
continuous phase transition located at a density close to the inverse of 
the location of the minimum of the respective pair potential ($r_m^{-1}$ = 
0.337 \AA$^{-1}$ vs $\lambda_s$ = 0.358 \AA$^{-1}$ for helium, and 
$r_m^{-1}$ = 0.291 \AA$^{-1}$ vs $\lambda_s$ = 0.309, 0.315 \AA$^{-1}$ for
molecular hydrogen).  

Inside the nanotube (T), the energies are much more negative that in 1D due
to the strong attraction of the carbon substrate: the binding energy of a
single H$_2$ molecule in the tube is $E_b = -1539.87 \pm  0.11$ K.  Looking at
the T-energy results contained in Table I one realizes that also in this 
case a transition occurs at a density very close to the
1D one. It is remarkable that both in 1D and T, H$_2$  remains liquid 
below the liquid-solid transition density, and thus a homogeneous liquid
phase at zero pressure is predicted. That result contrasts with the
theoretically and experimentally well established solid phase in 3D
\cite{sil2} and the
2D solid phase predicted by a PIMC calculation \cite{wag}.

The equations of state of liquid H$_2$ near the equilibrium density for 
both the 1D and T systems are shown in Fig. 1. In order to make the energy
scales compatible we have subtracted the single binding energy $E_b$ to the
T results. The lines in the figure correspond to the third-degree polynomial
fits in the form

\begin{equation}
\frac{E}{N} = e_0 +  A \left( \frac{ \lambda - \lambda_0 }{\lambda_0} \right)^2 +
B \left( \frac{ \lambda - \lambda_0 }{\lambda_0} \right)^3 \ .
\end{equation}

The best set of parameters $e_0$, $\lambda_0$, $A$ and $B$ are reported
in Table II. The equilibrium densities in both systems are the same considering
their respective uncertainties but the binding energy $e_0 = e(\lambda_0)$
is larger when H$_2$ is inside the nanotube. The 
difference between the 3D geometry (T) and the idealized one (1D) can be
quantified  by
means of the adimensional parameter
\begin{equation}
\Delta^{\rm T} =  \frac{(E^{\rm T} - E_b^{\rm T})- E_{\rm 1D}} 
{(E^{\rm T} -  E_b^{\rm T})}
\end{equation}
In the present system, around $\lambda_0$, $\Delta^{\rm T}$ = 3.5 \%  which
emphasizes the proximity between the real system and the idealized one. 
It is worth noting that in $^4$He inside the same nanotube we obtained 
$\Delta^{\rm T}=90$ \% \cite{gor}, and therefore H$_2$ seems a better candidate to
experimentally achieve a 1D condensed phase. That significant difference 
between helium and hydrogen may be understood taking into account that the 
difference between the hard-core size of the C-He interaction 
($\sigma_{C-He}$ = 2.74 \AA) and the C-H$_2$ one ($\sigma_{C-H_2}$ = 2.97
\AA) is magnified in a (5,5) tube because its small radius
($R$ = 3.42 \AA).

In Fig. 2, the density dependence of the pressure for both the 1D and T 
systems is reported from equilibrium up to the liquid-solid transition
density. As a matter of comparison, the same results  for $^4$He are also
plotted. Both in H$_2$ and  $^4$He the pressure increases faster in the
1D geometry ($P_{\lambda}$) than in the tube (P) due the transverse degree of
freedom that particles have in the latter case (notice the proportionality
between the scales of $P$ and $P_{\lambda}$ in Fig. 2, $P_{\lambda}/P = \pi
R^2$). At a given density $\lambda$, the difference between the T and 1D
pressures is smaller in H$_2$ than in $^4$He. For example, at the respective
transition densities that difference is more than one and a half times larger in helium than 
in hydrogen. Therefore, the one-dimensionality of H$_2$ inside the nanotube
is well maintained in all the liquid regime in contrast with  
$^4$He, in which the departure from such an idealized model already appears
around the equilibrium density and increases significantly with $\lambda$.  
Also apparent form Fig. 2 is a much smaller compressibility in H$_2$ than
in $^4$He. In the 1D geometry at $\lambda$ = $\lambda_0$ the velocity of
sound in H$_2$ is $c$ = 736.1 $\pm$ 0.2 m/sec to be compared with  
$c$ = 7.98 $\pm$ 0.07 m/sec in $^4$He at $\lambda$ = $\lambda_0$($^4$He) =
0.062 \AA$^{-1}$. The velocity of sound drops to zero at the spinodal point
that, according to the equation of state (Table II), is located at densities
$0.210 \pm 0.001$ \AA$^{-1}$ and $0.209 \pm 0.001$ \AA$^{-1}$ for the 
1D and  T systems, respectively.

The spatial structure of molecular hydrogen in the 1D array and inside the
nanotube has also been analyzed by means of the two-body radial distribution
function, $g(z)$, and its Fourier transform, the static structure factor
$S(k)$. 
In Fig. 3, 1D results for $g(z)$ are reported at both the equilibrium 
density for the liquid phase, and in the solid-liquid transition region
($\lambda$ = 0.312 \AA$^{-1}$). The corresponding results for H$_2$ inside the 
nanotube are indistinguishable from the 1D results in the scale shown in 
Fig. 3. The solid $g(z)$ shows a strongly localized order around the equally
spaced $z$-sites that decreases very slowly when $z$ increases. The result
for $g(z)$ at $\lambda$ =  $\lambda_0$ manifests the nature of a dense fluid
with an appreciable structure that decreases slowly and  
shows residual ordering up to large $z$ distances. A comparison with
previous DMC results on helium at the same density shows that the height 
of the $^4$He peaks is approximately an 80 \% of the corresponding ones 
in H$_2$, but with a $z$-ordering that also survives up
to large distances; the additional 20 \% comes from the sizeable  difference
between the well depths of the respective interaction potentials.  

The significant differences in structure between the low and high density 
regimes are reflected even more clearly in the static structure factors.
In Fig. 4, results for $S(k)$ corresponding to the same densities reported
in Fig. 3 are shown. At high density, a characteristic result for a
solid phase is obtained with a regular $k$-spacing according to the only
periodicity allowed by the 1D geometry. At $\lambda$ =  $\lambda_0$, 
$S(k)$ shows a first peak reflecting the  
localization observed in $g(z)$ (Fig. 3), and a subsequent very smoothed
maximum as expected in a homogeneous liquid phase. 

In conclusion, we have studied the zero-temperature equation of state of 
molecular hydrogen in a 1D geometry and inside a narrow nanotube by means
of the diffusion Monte Carlo method. The 1D calculation predicts the 
existence of a self-bound system with a binding energy of -4.8 K and a 
quasi-continuous liquid-solid transition at high density. The comparison
with a real system, hydrogen in a nanotube, points to a close
proximity between its properties and the ones of the 1D limit. 
The prediction of a liquid H$_2$ phase inside a (5,5) carbon nanotube 
is one of the main conclusions of the present work. The high one-dimensionality
of this system would preclude a superfluid behavior but the use of wider nanotubes
can provide a proper setup to observe the so-long desired superfluidity
in molecular hydrogen. We expect the present work could encourage 
experimentalists to explore such a intriguing possibility.

This research has been partially supported by DGES (Spain) Grants N$^0$
PB96-0170-C03-02 and PB98-0922, and DGR (Catalunya) Grant N$^0$
1999SGR-00146. M. C. G. thanks the Spanish Ministry of Education and Culture 
(MEC) for a postgraduate contract. We also acknowledge the supercomputer facilities
provided by the CEPBA.

\begin{figure}
\begin{center}
\epsfxsize=5.5cm  \epsfbox{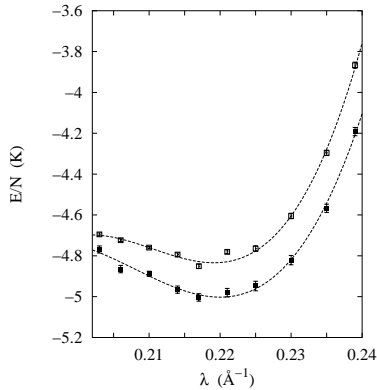}
\caption{Energy per particle of H$_2$ as a function of the linear density. Open
squares are the 1D results, and filled squares are the T energies having
subtracted the binding energy of a single molecule $E_b$. The lines are the
result of the polynomial fit (Eq. 4) with the optimal parameters reported
in Table II.}
\end{center}                                      
\end{figure}                            

\begin{figure}
\begin{center}
\epsfxsize=5.5cm  \epsfbox{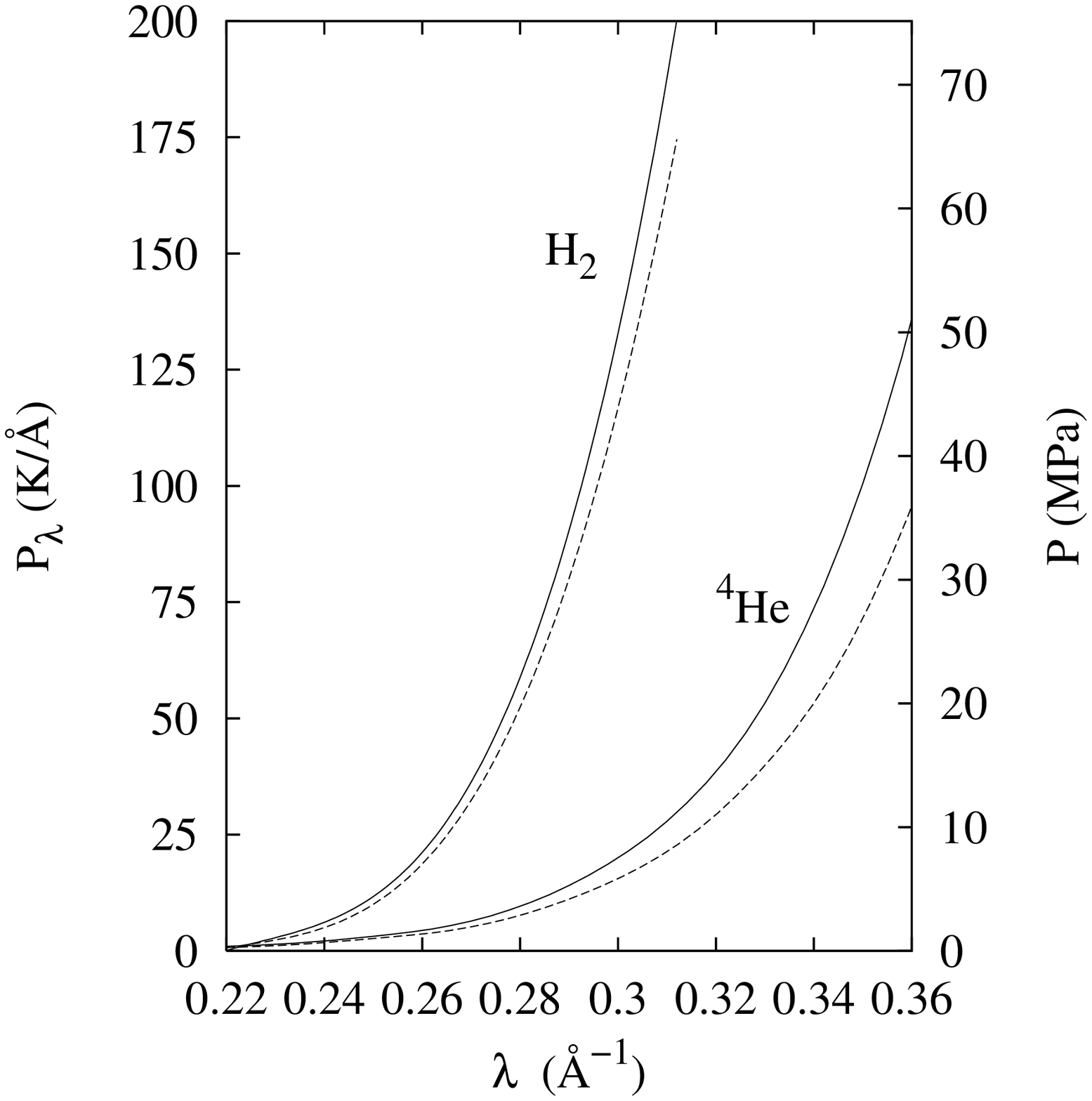}
\caption{1D ($P_{\lambda}$, solid line) and T ($P$, dashed line)  
pressures for H$_2$ and $^4$He as a function of the linear density.}
\end{center}                                      
\end{figure}              

\begin{figure}
\begin{center}
\epsfxsize=5.5cm  \epsfbox{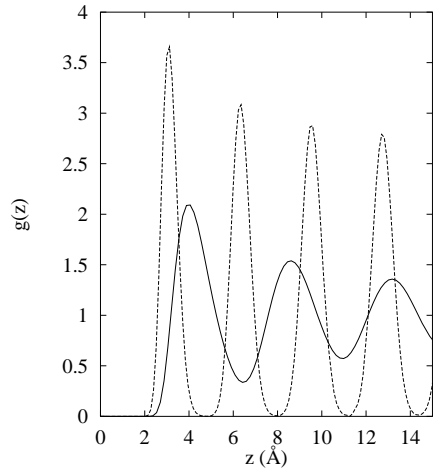}
\caption{Two-body radial distribution function for 1D H$_2$ at
equilibrium (solid line) and at the liquid-solid transition density (dashed
line).} 
\end{center}                                      
\end{figure}              

\begin{figure}
\begin{center}
\epsfxsize=5.5cm  \epsfbox{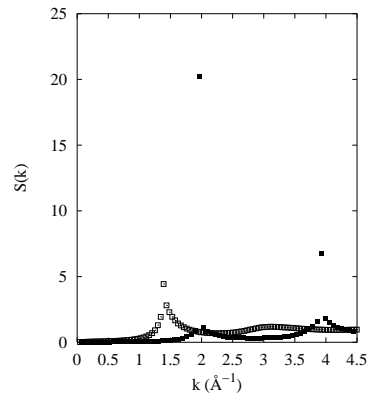}
\caption{Static structure function for 1D H$_2$ near
equilibrium (open symbols) and at the liquid-solid transition density
(filled symbols).} 
\end{center}                                      
\end{figure}              

\begin{onecolumn}
\begin{table}
\begin{tabular}{lcccc}
$\lambda$ (\AA$^{-1}$) &  $E/N$ (1D, $a$ = 0)  & $E/N$ (1D, $a \not=$ 0)
& $E/N$ (T, $a$ = 0) & $E/N$ (T, $a \not=$ 0)
\\ \hline
0.329  &   98.083 $\pm$    0.034  &   97.963  $\pm$  0.016 &   -1453.99 $\pm$ 0.06 & -1454.69 $\pm$ 0.04 \\
0.320  &   72.567 $\pm$    0.013  &   72.523  $\pm$  0.007 &   -1476.74 $\pm$ 0.05 & -1476.88 $\pm$ 0.01 \\
0.312  &   53.264 $\pm$    0.010  &   53.227  $\pm$  0.010 &   -1493.790 $\pm$ 0.019 &  -1493.720 $\pm$ 0.002 \\
0.304  &   38.581 $\pm$    0.018  &   38.636  $\pm$  0.014 &   -1506.570 $\pm$ 0.03 & -1506.540 $\pm$ 0.011  \\
0.290  &   19.203 $\pm$    0.010  &   19.260  $\pm$  0.003 &   -1523.730 $\pm$ 0.017 & -1523.600 $\pm$ 0.02 \\
\end{tabular}
\caption{Energies per particle in K at high linear densities $\lambda$ for
1D and T H$_2$ systems. $a$ = 0 and $a \not=$ 0 correspond to the liquid
and solid phases, respectively.}
\end{table}
\end{onecolumn}

\begin{table}
\begin{tabular}{lcc}
Parameter   &  1D H$_2$  & H$_2$ in a tube \\ \hline
$\lambda_0$ (\AA$^{-1})$ & 0.2191 $\pm$ 0.0004 & 0.2200 $\pm$ 0.0006 \\
$e_0$ (K)               & -4.834 $\pm$ 0.007 & -1544.880 $\pm$ 0.016 \\
$A$   (K)               & 65.7 $\pm$ 3.6 & 69.7 $\pm$ 4.9 \\
$B$   (K)               & 556.0  $\pm$ 46.6 & 429.7 $\pm$ 73.9 \\
$\chi^2/\nu$            &         1.98        &    1.5           \\
\end{tabular}
\caption{Parameters of the equation of state (Eq. 4) for the two systems 
studied.}
\end{table}

\end{document}